\documentclass[aps,prl,twocolumn,nofootinbib,superscriptaddress,floatfix]{revtex4-1}  
\usepackage{graphicx}  
\usepackage{amsmath}
\usepackage{amssymb}   
\usepackage{color}
\usepackage{hyperref}
\usepackage{ulem}

\def\gsim{\;\rlap{\lower 2.5pt \hbox{$\sim$}}\raise 1.5pt\hbox{$>$}\;}
\def\lsim{\;\rlap{\lower 2.5pt \hbox{$\sim$}}\raise 1.5pt\hbox{$<$}\;}

\begin{document}

\title{Cosmic explosions, life in the Universe and the Cosmological Constant}
\author{Tsvi Piran}
  \email{tsvi.piran@mail.huji.ac.il}
 \affiliation{Racah Institute of Physics, The Hebrew University, Jerusalem 91904, Israel} 

\author{Raul Jimenez}
  \email{raul.jimenez@icc.ub.edu}
 \affiliation{ICREA \& ICC, University of Barcelona, Marti i Franques 1, Barcelona 08028, Spain.} 
 \affiliation{Institute for Applied Computational Science, Harvard University, MA 02138, USA.}
 
\author{Antonio J. Cuesta}
  \email{ajcuesta@icc.ub.edu}
 \affiliation{ICC, University of Barcelona, Marti i Franques 1, Barcelona 08028, Spain.} 
 
\author{Fergus Simpson}
  \email{fergus2@icc.ub.edu}
 \affiliation{ICC, University of Barcelona, Marti i Franques 1, Barcelona 08028, Spain.} 

\author{Licia Verde}
\email{liciaverde@icc.ub.edu}
 \affiliation{ICREA \& ICC, University of Barcelona, Marti i Franques 1, Barcelona 08028, Spain.} 
 \affiliation{Institute of Theoretical Astrophysics, University of Oslo, Oslo 0315, Norway.}

\date{\today}

\begin{abstract}
Gamma-Ray Bursts  (GRBs) are copious sources of gamma-rays whose interaction with a planetary atmosphere can pose a threat to complex life. Using recent determinations of their rate and  probability of causing massive extinction, we explore what type of universes are most likely to harbour advanced forms of life.  We use cosmological N-body simulations to determine at what time and for what value of the cosmological constant ($\Lambda$) the chances of life being unaffected by cosmic explosions are maximised. Life survival to GRBs favours $\Lambda$-dominated universes. Within a $\Lambda$CDM cosmology, the likelihood of life survival to GRBs is governed by the value of $\Lambda$ and the age of the Universe.  We find that we seem to live in a favorable point in this parameter  space that minimises the exposure to cosmic explosions, yet maximises the number of main sequence (hydrogen-burning) stars around which advanced life forms can exist.
\end{abstract}

\pacs{}
\maketitle

Why the value of the cosmological constant $\Lambda$ is neither zero nor of the order of the Planck density (the Planck mass to the fourth power, $M_{\rm Pl}^4$) remains one of the deepest mysteries of nature. Refs~\cite{barrow, Weinberg} have argued that in order for observers (and thus cosmic structure) to exist, the value of $\Lambda$ could not be larger than $10^{-120} M_{\rm Pl}^4$. This was the first indication that the value of $\Lambda$ could not be arbitrary and that requiring the existence of observers bounded $\Lambda$  from above. Subsequent works  
(e.g., ~\cite{Efstathiou,TegmarkRees,Garriga99,Garriga00,Peacock}) have firmed up this argument.   
However, to this date no argument has been given to provide a lower bound to  $\Lambda$; in particular, it is not clear why $\Lambda$ does not simply vanish\footnote{A lower bound to $\Lambda$ based on the stability of atoms\cite{BarrowSandvik}, applies only in the  distant future for spatially flat universes.}. Interestingly, the necessity to avoid massive life extinction events by GRBs can shed a new light on this issue.

GRBs are potentially catastrophic events for biological organisms\footnote{In what follows when we refer to life we consider biological organisms which are sufficiently complex to act as observers. Our considerations will be for Earth-like planets where the UV protection provided by the atmosphere is due to an Ozone layer.}. In particular,  copious flux of $\gamma$-ray photons with energies above $10-100$ keV could destroy the ozone layer of a --habitable, Earth-like--  planet,  exposing living organisms to damaging UV radiation and compromising its habitability. 
This has led to the suggestion  \citep{Thorsett95,Scalo+02,Melott04} that galactic GRBs have been responsible for some  mass extinction events on Earth. Yet the rate or energy of nearby GRBs were not sufficient to avoid the emergence of  observers. However, such GRBs take place more frequently at the inner parts of the Milky Way and may cause a serious problem for development of life there \cite{PJ14}. On Earth and in general, in the outskirts of large galaxies, the most luminous GRBs, the ones around the knee  of the luminosity function see Fig.~3 in \cite{PJ14},  pose the greatest threat for the development of complex organisms \footnote{A small fraction of mass extinction from  short-GRBs  \cite{Melott06} won't change our results, as it would just add a baseline that equally penalises all values of $\Lambda$.}as they could cause catastrophic damage even if located in a sufficiently nearby satellite galaxy.  

The rate of GRBs within a given galaxy depends on the metallicity: most GRBs take place  where  metallicity $< 0.3$ solar and the stellar mass of the host galaxy is above $5 \times 10^7$ M$_{\odot}$   \cite{Savaglio,JP13}. Such low-metallicity environments are rare within the Milky Way and the $\sim 50$kpc region around it. 
Small mass, low metallicity,  Magellanic Cloud (SMC and LMC)-type galaxies are the typical host of GRBs and  thus the most likely location for potentially damaging nearby GRBs. We explore next the rate of such catastrophic extra-galactic events.

The observed global GRB rate is $10^9$ GRB/Gpc$^{3}$/Gyr \cite{WP}\footnote{We use here the rate of GRBs beamed towards a given observer. The overall rate is larger by a factor corresponding to the beaming. However, this is not relevant for this work}. GRB hosts have stellar masses between $5 \times 10^{7}$ and $10^{10} M_{\odot}$ \cite{Savaglio,JP13}. Integrating the stellar mass function of \cite{Panter} in this mass range we find a stellar density of $10^{16} M_{\odot}/Gpc^3$, yielding a rate of $10^{-7}$ GRB/$M_{\odot}$/Gyr. This rate depends only on stellar physics and thus is independent of cosmology.  Integrating (out to $200$ kpc from the center) the dark matter profile from CDM simulations \cite{Acquarius}, the dark matter mass in satellites is 20\% of the total halo mass ($2 \times 10^{12}$) of the Milky Way. Since  $\sim1$\% of this mass is in stars \cite{Fukugita} we obtain a stellar mass of $4 \times 10^{9}$ in the satellites\footnote{This number is in excellent agreement with direct integration of the observed stellar mass in Milky Way satellites\ \cite{James11}}. Thus, we expect $400$ GRB/Gyr in satellites. Using the observed GRB luminosity function \cite{WP} assuming an effective duration of $10s$, we expect: $280$ (70\%)  $>10^{52}$ erg; $72$ (18\%)  $>10^{53.5}$ erg; $4$(1\%)  $>10^{54.5}$ and $0.5-1$(0.15\%)  with energy of $\sim 10^{55} erg$. Ref~\cite{PJ14} has shown that most likely  there has been one GRB during the last Gyr with a fluence on Earth of $100 kJ/m^2$; this fluence is the value found by \cite{thomas,Melott} for massive life extinction to take place. This event is believed to have caused the Orvidician extinction\cite{Ordo}, which wiped out 85\% of all species present on Earth at the time. Following \cite{thomas,Melott}
we take a  fluence of $100 kJ/m^2$ to be the threshold: higher fluence would have catastrophic consequences for having observers\footnote{The amount of ozone depletion and DNA damage scales slowly with fluence: they are reduced by factors of 2 and 2.5 respectively by  reducing  fluence from 100  to 10  kJ/m$^2$ \cite{thomas06}.}. Then the equivalent damaging distance from the center of the galaxy for a planet at 10 kpc from the center is 
17kpc for $10^{53.5}$ erg, 27 kpc  for $10^{54.5}$ and 50 kpc for $10^{55}$ erg. 
This implies, conservatively, that a region of $20kpc$  (from the center of the host halo)  should be devoid of GRB-hosting sub-halos for  harbouring planets suitable to  support observers. If there are no satellites in this region then there will be no damaging GRBs (of fluence $100kJ/m^2$) with a  rate much higher\footnote{About 8\% of the satellites are within a radius of 20kpc for a $\Lambda=0$ cosmology. Regardless of the cosmology, almost all satellites are within a radius of $50-100$ kpc.} than that on Earth. Since there is a non-zero probability of a $\geq 10^{55} erg$ GRB/Gyr in the satellites which yields a lethal $100kJ/m^2$  fluence at a  distance of $50$kpc, we also discuss this case.
 
The existence of many nearby satellites will have an additional effect. Many of these satellites in-fall into the main galaxy, bringing low metallicity material and  triggering further star formation that will increase  GRB production. We do not consider this enhanced rate in this work but it clearly makes development of intelligent life in cosmologies with numerous nearby satellites even more difficult than considered here. 

Any inference of cosmological parameters ought to take into account the selection effects which have led us to observe the Universe from this particular vantage point  \cite{Fergus}. Of particular interest is the value of $\Lambda$, which can govern the growth of regions around large galaxies devoid of LMC-type satellites. The accelerated expansion induced by a cosmological constant  slows the growth of cosmic structures, and increases the mean inter-galaxy separation. This  reduces the number of  nearby satellites likely to host catastrophic GRBs. Below we quantify this effect.

 \begin{figure}
\includegraphics[width=1.05\columnwidth]{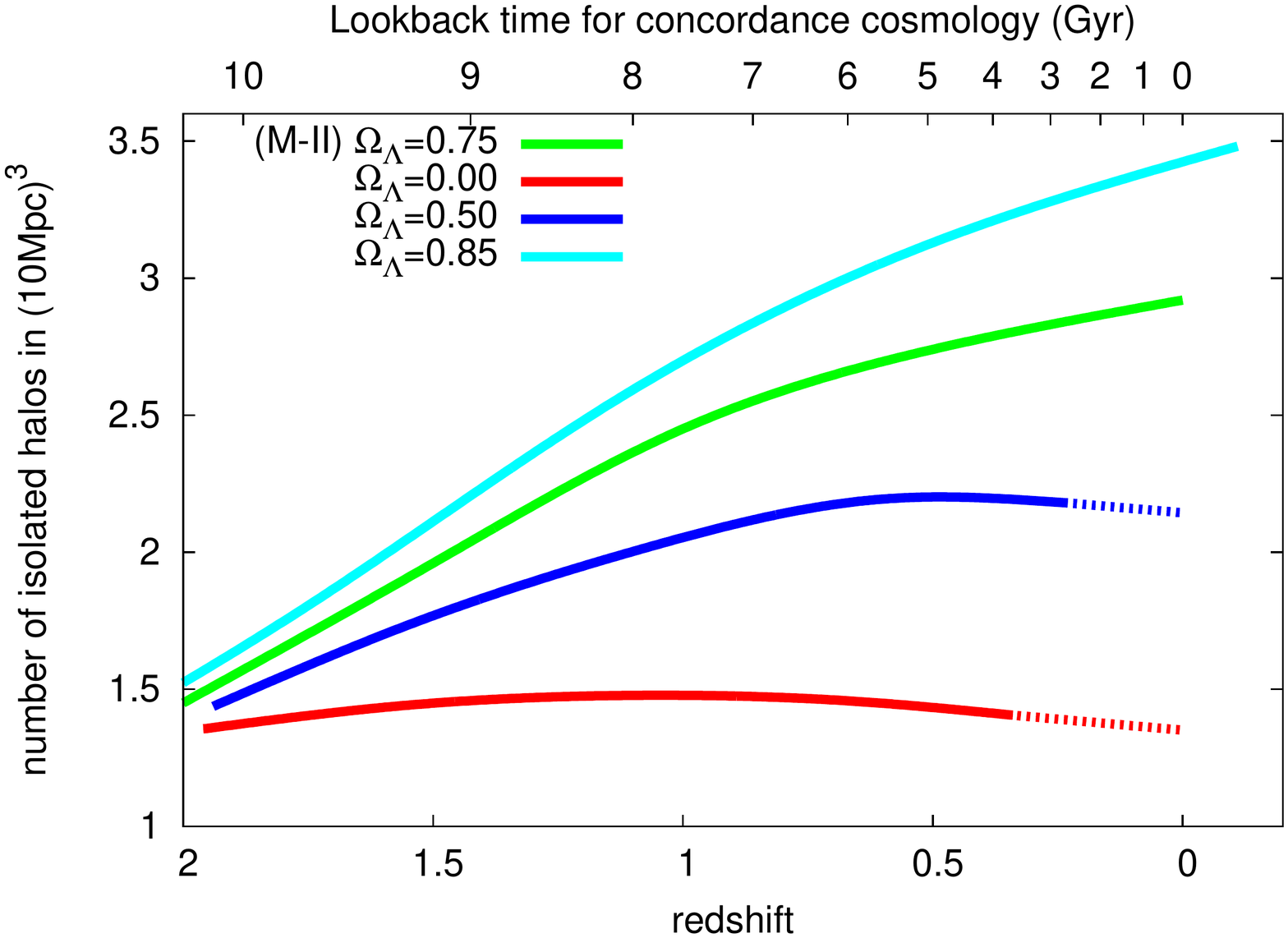} 
\includegraphics[width=1.05\columnwidth]{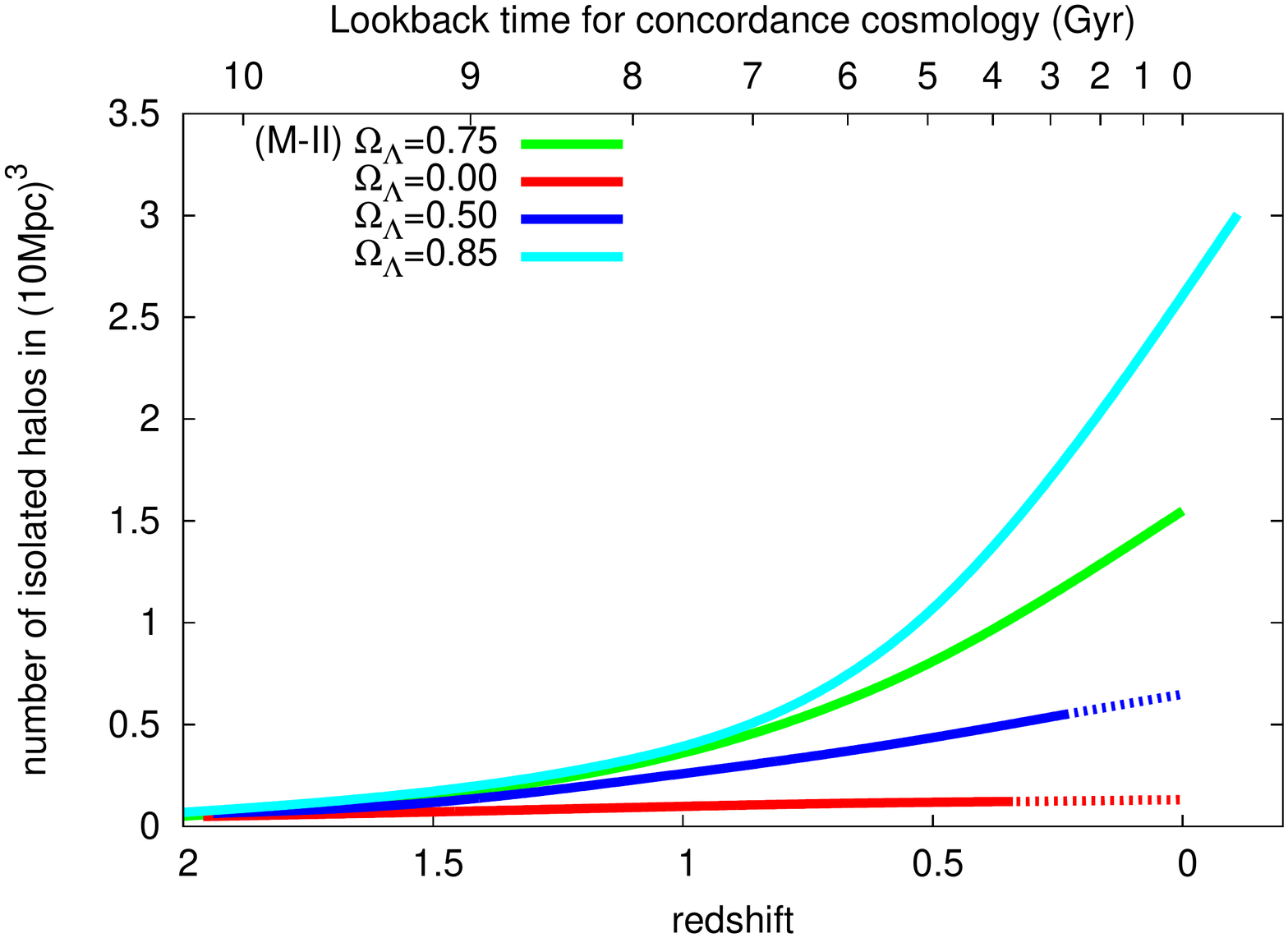}
\caption{\label{fig:lcdm}  Number of isolated halos as a function of redshift that are protected from the damaging effect on life of GRBs; we show four cosmologies and two isolated radii of satellites: 20 kpc (upper panel) and 50 kpc (lower panel). To guide the eye, dashed lines show extrapolations to $z=0$ due to limitations in re-mapping the Millenium-II simulation as this would have needed it to be run into the future. The number of life-protected regions is significantly larger in the $\Lambda-$dominated cosmologies. This is due to the effect of $\Lambda$ at late times at clearing and smoothing virialised regions. Also remarkable is the fact that only below $z \approx 1$ the number of isolated halos grows significantly. This is about $7$ Gyr ago, not too dissimilar from the age of Earth.}
\end{figure}

Using $N$-body simulations, we search for  halos with dark matter masses  $10^{11.5} < M/M_{\odot} < 10^{12}$ that have no satellites of dark matter mass $ > 2 \times 10^8 M_{\odot}$ within a radius of 20 (50) kpc (proper); we refer to these halos as isolated. In particular, we use the \textsc{Millennium-II}   \cite{MII} publicly available dark-matter-only simulation \footnote{http://www.mpa-garching.mpg.de/galform/millennium-II/} through the Millennium Database portal created by the Virgo Consortium  \cite{Lemson}. We  search for these isolated halos over all 37 available snapshots from $z=2$ to $z=0$ to determine the redshift evolution.
We determine the number of subhalos $N_{\rm sh}$ inside the search radius for each host halo (identified with a Friends-of-Friends ID number), which can be either empty ($N_{\rm sh}=0$, isolated host halo) or not empty ($N_{\rm sh}>0$, non-isolated host).  

The \textsc{Millenium-II} simulation is performed for a flat $\Lambda$CDM cosmology with parameters\footnote{Unless otherwise stated, cosmological parameters are at $z=0$.}: $\Omega_m = 0.25$, $\Omega_b = 0.045$, $\Omega_{\Lambda}=0.75$, $h = 0.73$, $\sigma_8 = 0.9$, $n_s =1$. 
The size of the box is $L=100h^{-1}$Mpc (comoving) on a side. The spatial resolution is $1h^{-1}$kpc, so the search radius for subhalos is not affected by resolution effects. The mass resolution of the simulation is $6.89 \times 10^6 h^{-1}M_{\odot}$, so effectively our threshold on the mass of the satellite halo is verified by \textit{all} subhalos resolved in the simulation, regardless of their mass. The host halo is required to have  30,000 to 100,000 particles, which corresponds to the halo mass bin stated above. 

No public N-body simulations with the required mass resolution exist for other models than the ``vanilla" $\Lambda$CDM. 
Fortunately, numerical simulations of one cosmology can be re-mapped into a different one accurately both for the dark matter field and the corresponding halos \cite{Angulo,MeadPeacock}. Using this algorithm we have re-mapped the results of the Millennium-II $\Lambda$CDM  simulation to cosmologies with other values of  $\Lambda$. Note that by keeping the geometry fixed ($\Omega_m+\Omega_{\Lambda}=1$) and the  early universe quantities (such as the physical matter density  $\Omega_m h^2$, the amplitude of primordial perturbations, and the baryon fraction) fixed   the current value of $\Omega_{\Lambda}$ specifies the cosmology.

It is challenging to keep track of changes of all other cosmological parameters and it is clear that other changes could mimic some of the effects that we emphasise here, for example significant amount of massive neutrinos would also suppress the growth of structure. 
However, it has been shown \cite{Tegmark05} that even modest modifications to a range of cosmological parameters leads to adverse consequences for the abundance of life. As such, we focus on changing only the value of the cosmological constant, while keeping the conditions of the early universe fixed.

\begin{figure}
\includegraphics[angle=0,width=\columnwidth]{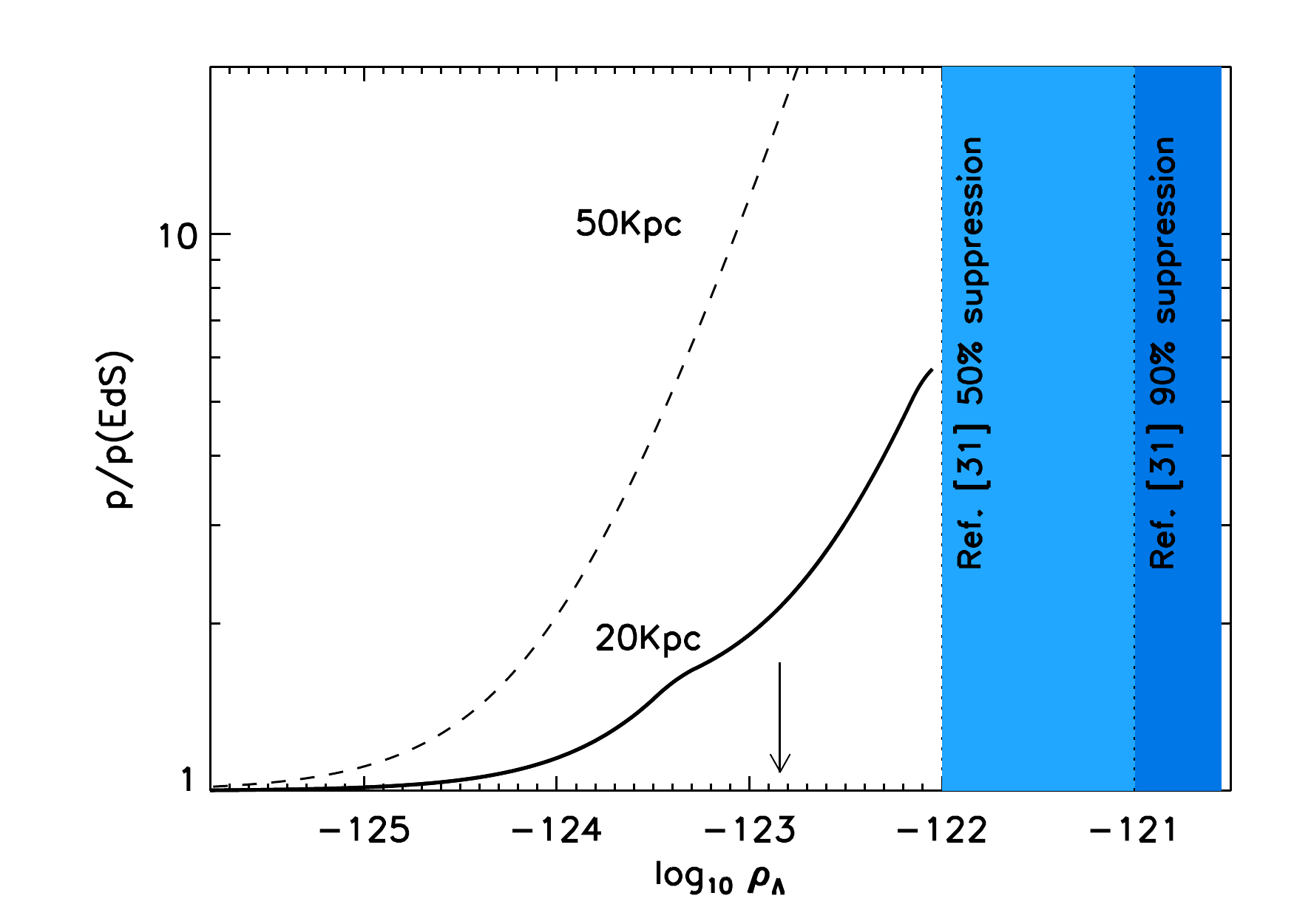}
\caption{\label{fig:pdf} The probability distribution function for $\rho_\Lambda$  normalised to the probability for the Einstein-de-Sitter (EdS, $\Lambda=0$) case. The solid (dashed)  line  corresponds  to exclusion regions of $20$ ($50$) kpc and result from this work. 
For high $\Lambda$ values, the exponential suppression of Ref. \cite{Tegmark05}  takes over. This is indicated by the shaded regions. The arrow shows the value of $\rho_\Lambda$ for the concordance $\Lambda$CDM model. Our result suppresses the probability of low values of  $\Lambda$ previously allowed, and in fact favoured, by the argument in  \cite{barrow,Weinberg}.}
\end{figure}

Fig.~\ref{fig:lcdm} shows the number of isolated halos per comoving $1000$ Mpc$^3$  for four relevant flat cosmologies spanning the range $0\le \rho_{\Lambda}/M_{\rm Pl}^4\le 2.7\times 10^{-123}$: {\it 1)} $\Omega_m=0.15, \Omega_{\Lambda}=0.85, \rho_{\Lambda}/M_{\rm Pl}^4 = 2.7\times 10^{-123}$, {\it 2)} (fiducial $\Lambda$CDM) 
 $\Omega_m=0.25, \Omega_{\Lambda}=0.75, \rho_{\Lambda}/M_{\rm Pl}^4 = 1.4\times 10^{-123}$, {\it 3)} $\Omega_m = 0.5$, $\Omega_{\Lambda} = 0.5, \rho_{\Lambda}/M_{\rm Pl}^4 = 4.5\times 10^{-124}$ and {\it 4)} (Einstein de Sitter) $\Omega_m=1$, $\Omega_{\Lambda} = 0, \rho_{\Lambda}=0$ . In the $\Lambda$ dominated models, only fairly recently, on a cosmological time scale, the number of GRB-protected halos grew significantly. In fact, a large number\footnote{For a MW-like halo in the concordance $\Lambda$CDM model, $1/3$ of the halos fulfil the 50 kp isolation criterion, so our MW is not a very special halo} of isolated halos appear only below $z \approx 1.5$. This corresponds to a lookback-time of about $7$ Gyr, not dissimilar from the age of Earth. As structure formation proceeds faster in the $\Lambda=0$ universe, the amount of sub-structure grows faster and earlier: the number of isolated halos is much smaller than in the case of a $\Lambda$ dominated universe. 

In order for a particular galaxy to harbour life, it must reside within a habitable region of the parameter space.
The halo mass should be in the range $10^{11.5}  < M/M_{\odot} < 10^{12}$ as to give rise to a large galaxy like the Milky Way so that it has significant outer regions in which the GRB rate is low but there are sufficient number of stars. Additionally the galaxy should undergo sufficient chemical evolution so that  its average metallicity is relatively large thereby reducing the GRB rate.
Furthermore,  the halo should be  young enough that stars with mass $M_{\rm min}$ have not left the main sequence. For  $M>0.5M_{\odot}$ stars  this corresponds to  ($\lesssim 50$ Gyr), for $M_{\rm min}>0.7M_{\odot}$ this corresponds to $\lesssim 20$ Gyr. For example for the model with $\Omega_{\Lambda} = 0.5$ the age of the Universe is $\sim 20$ Gyr, by the time $\Omega_{\Lambda}$ takes over the expansion, the universe is already too old and out of the habitable epoch for all main sequence stars above $M_{min}=0.7 M_{\odot}$. 
For $M_{\min}=0.5 M_{\odot}$  we estimate this  to  correspond instead to $\Omega_{\Lambda}=0.2$ ($\rho_{\Lambda}/M_{\rm Pl}^4 = 1.2\times 10^{-124}$). 

One can imagine  waiting for a long time for the GRB rate to be sufficiently low.  The current decrease of the GRB rate with time \cite{WP} at $z <1$ is much flatter than the decrease in  star formation rate (see their Fig.~9); extrapolating this rate implies that the star formation will be exhausted  sooner than GRBs. By $z \sim =-0.4$-- when flat universes with  a cosmological constant that is 0.3-0.6 of that in the vanilla model become similar to the present  Universe with $\Omega_\Lambda =0.75$, the GRB rate will be down by a factor $2$, while the star formation rate will be down by a factor $10$. Since the decrease in  star formation rate is the dominant effect, for simplicity in what follows we  assume that the GRB rate remains constant. This approximation does not change  qualitatively our argument.

In the range of $\rho_{\Lambda}$ that we have explored, and for $M_{\rm min}=0.7 M_{\odot}$, we find that the number density of habitable and isolated halos  can be approximated by\footnote{For  50 kpc radius  $\eta= 0.1$ and $\kappa=1.1\times 10^{123}$ for  $\rho_{\Lambda}$ in $M_{\rm PL}^4$. For 20 kpc radius we find that a broken power law is a slightly better fit:  $\kappa_1=0.72 \times 10^{123}$, $\eta_1=0.52$; $\kappa_2=0.24\times 10^{123}$, $\eta_2=0.72$ respectively at low and high values of  $\rho_{\Lambda}$.}
\begin{equation}
I(\rho_{\Lambda}) \sim \kappa \rho_{\Lambda}+\eta \,.
\label{eq:ILambda}
\end{equation}
The coefficients of the fit may change for different values of  $M_{\rm min}$ or different radii but the qualitative behaviour remains  with a sharp  decrease for $\Omega_{\Lambda}<0.2$  $(\rho_{\Lambda}< 1.2\times 10^{-124})$.
For larger values of $\Lambda$ this relation  must flatten  since  virtually  all MW-size halos are isolated  for $\Omega_{\Lambda} > 0.85$. 
We also note that, always in the range of  $\rho_{\Lambda}$  explored, the number of MW-sized halos in the same volume  is roughly constant, implying that $I(\rho_{\Lambda})$ is roughly proportional to the fraction of isolated halos (for larger values of $\Lambda$ when this fraction reaches unity  it is assumed  to become a constant). If we assume that, for the values of $\rho_{\Lambda}$ we considered  so far, $I(\rho_{\Lambda})$ is proportional to the probability $p(I|\rho_{\Lambda})$, to infer the posterior probability for having a cosmological constant $p(\Lambda | I) \propto   p(\Lambda)   p(I | \Lambda) \,$  we must define a prior distribution $p(\Lambda)$. Previous studies take a flat prior on $\rho_\Lambda$ based on the premise that its value may arise from the cancellation of much larger terms.  However given the large uncertainty regarding the nature of dark energy, this might not be the case. A more conservative approach is to select an uninformative prior such that $p(\Lambda) \propto 1/ \rho_{\Lambda}$ (see also \cite{Garriga00}). In this case one  shall not consider negative values of the cosmological constant, although they are likely to generate even more inhospitable consequences. 

The other mechanism by which $\Lambda$ influences the habitability of the Universe is the suppression of galaxy formation. This was studied in detail in \cite{Tegmark05} who used the fraction of baryons $F(\mu)$ residing in halos of mass $\mu \equiv \xi^2 M$, where  $\xi$ is  the matter density per photon, as a proxy for the habitability of the universe. They present the following prescription for the late time solution
\begin{equation}
F(\mu) = \text{erfc} \left[\frac{A \rho_\Lambda^{1/3} }{ \xi^{4/3} Q s(\mu) G_\infty} \right] \, .
\label{eq:uppercutoff}
\end{equation}
Here  $A$ and $G_\infty$ can be considered fixed quantities,  $A = 5.59$ and $G_{\infty}  = 1.43$; we take $s=28$ and $\xi^4 Q^3=10^{-124}$ following \cite{Tegmark05} (see e.g., their Fig 7.). It is clear from  Eq.~\ref{eq:uppercutoff} and  e.g., Fig. 7 of   \cite{Tegmark05}  that this imposes a  sharp suppression for $\rho_{\Lambda}\gtrsim 10^{-122}$ M$_{\rm Pl}^4$. 

In summary, 
\begin{align}
p(\Lambda | I) 
&\propto p(\Lambda) F(\Lambda) I(\Lambda)
\end{align}
where $F(\Lambda)$ gives the number of suitable MW-size halos and   describes the sharp upper cut-off  imposed by eq. \ref{eq:uppercutoff}, and $I(\Lambda)$ imposes a  much slower suppression towards 0.
In Fig.~\ref{fig:pdf} we sketch the probability distribution function for $\rho_\Lambda$, normalised to the Einstein-de-Sitter case, using the computed abundance of isolated Milky Way-size halos from Fig.~1. 
The low-end suppression is described by Eq.~\ref{eq:ILambda}. For high values of  $\rho_\Lambda$ the suppression discussed in   \cite{Tegmark05, barrow,Weinberg}, effectively due to the requirement of cosmic structure to form,  becomes relevant. It is not surprising then that our universe has a value of $\Lambda \approx 1$ ($\rho_\Lambda \sim 10^{-123}M_{\rm Pl}^4$). It is  instructive to compare to Fig.7 of \cite{Tegmark05} and notice that the allowed region for $\rho_{\Lambda}$ is now greatly reduced.

Since on average there is  {\it one} isolated region in a  patch of  10 Mpc radius (which, incidentally, is the mean inter galaxy cluster separation today), to ensure (at $\sim  3\sigma$) that there is at least one habitable galaxy in the observable universe,  the horizon size should be  at least 40 Mpc i.e. not more than $\sim 100$ times smaller than the current horizon size. Given our specifications for galactic habitability, both in terms of the required separations between galaxies, and the minimum age of the Universe which permits the formation of planets, a large universe is necessary for life to emerge. A loitering model with a finely tuned cosmological constant could satisfy the age requirement, but does not provide adequate inter-galactic spacing. 

If our location in the Universe, and potentially the multiverse, is preferentially selected by the absence of nearby, GRB-hosting small halos, then we will expect to find that most Milky-Way sized galaxies beyond the Local Group possess a heightened number of LMC-like satellites. To some extent this has already been observed \cite{cote}. The reduced number of observed Milky-Way satellites compared to predictions derived from simulations is often attributed to the inefficiency of star formation within low mass halos. Yet this puzzle may also be partially resolved by our proposed selection effect. 

Another interesting implication is that if the amplitude of fluctuations $Q$ is increased, then halos of a given mass form earlier, but they do so in a very crowded environment. In previous investigations $Q$ is one of the few cosmological parameters which could be enlarged by an order of magnitude without any clearly adverse effects (e.g., Fig.12 in \cite{Tegmark05}). Within the context of GRBs-limited habitability, there is likely to be much less freedom in this parameter. 

In summary, we have shown that  $\Lambda$ plays a crucial role at creating  habitable regions for galaxies in a habitable epoch. These considerations may be used to disfavour very low values for $\Lambda$. Negative values of $\Lambda$ will yield even more satellites and hence these arguments strongly disfavour such values.

\begin{acknowledgments}
We thank R. Angulo for help with  interpreting the  simulation outputs. This research was supported by the ISF I-Core center of excellence and by an Israel-China grant (TP),  by Mineco grant AYA2014-58747-P (RJ,LV), FP7-IDEAS-Phys.LSS 240117 (LV, AJC,FS) and the Spanish MINECO under project MDM-2014-0369 of ICCUB (Unidad de Excelencia `Mar{\'\i}a de Maeztu').

\end{acknowledgments}

\newcommand{\jcap}{Journal Cosmology and Astroparticle Physics} 
\newcommand{\apjl}{The Astrophysical Journal Letters}
\newcommand{\mnras}{MNRAS}
\newcommand{\aap}{Astronomy \& Astrophysics}
\newcommand{\araa}{ARA\&A}
\newcommand{\pasp}{PASP}

\end{document}